\documentstyle[12pt,psfig]{article}
\def\labell#1{\label{#1}}
\def\Tr{{\rm Tr}}

\def\D{{\rm D}_q}
\def\DD{{\rm D}_{q_2}}
\def\DDD{{\rm D}_{q_1}}

\begin{document}

{\flushright{\small NSF-ITP/97-123\\Imperial/TP/97-98/7\\PUPT-1735\\
hep-th/9711008\\}}

\begin{center}
{\Large  {\bf Boundary fields and renormalization group flow}}
\\
{\Large  {\bf in the two-matrix model}}

\vspace{0.2in}
Sean M.  Carroll
\vskip 0.2cm
{{\it Institute for Theoretical Physics, University of
California, \\
Santa Barbara, California 93106, USA}
\\
\small E-mail: {\tt carroll@itp.ucsb.edu}}

\vspace{0.2in}
Miguel E.  Ortiz
\vskip 0.2cm
{{\it Blackett Laboratory, Imperial College of Science, Technology
and Medicine,\\ 
Prince Consort Road, London SW7 2BZ, UK}
\\
\small E-mail: {\tt m.ortiz@ic.ac.uk}}

\vspace{0.2in}
{Washington Taylor IV}
\vskip 0.2cm
{{\it Department of Physics, Princeton University,
Princeton, New Jersey 08544, USA}
\\
\small E-mail: {\tt wati@princeton.edu}}
\end{center}

\vskip 1truecm

\begin{abstract}
We analyze the Ising model on a random surface with a boundary magnetic 
field using matrix model techniques.  We are able to exactly calculate
the disk amplitude, boundary magnetization and bulk magnetization 
in the presence of a boundary field.  The results of these calculations
can be interpreted in terms of renormalization group flow induced by the
boundary operator.  In the continuum limit this RG flow corresponds to
the flow from non-conformal to conformal boundary conditions which has
recently been studied in flat space theories.
\end{abstract}

\medskip

\newpage

\renewcommand{\baselinestretch}{1}
\baselineskip 18pt

\section{Introduction}
\label{sec:introduction}

Simple statistical mechanical lattice models, such as the Ising model,
have been used for many years to gain insight into the behavior of a
wide range of physical systems.  The great utility of such simple
models arises from the fact that in many cases they are exactly
solvable.  It is a remarkable fact that some
questions which seem analytically intractable for models on a regular
lattice yield exact solutions when the underlying lattice is itself
taken to be a random element of a larger statistical ensemble.  An
example of such a situation is the Ising model with a bulk magnetic
field.  Although this model has not been analytically solved on a
fixed lattice, it is possible to exactly compute the partition
function and magnetization of the model on a random lattice
\cite{kazakov}.

In this paper, we consider the Ising model on a random lattice in the
presence of a magnetic field on the boundary, rather than in
the bulk.  Again, this corresponds to a
problem which does not seem to be analytically solvable on a
fixed lattice.  Summing over random lattices, however, we find that the
partition function and magnetizations can be calculated exactly.

The Ising model on a random two-dimensional lattice can be described
in terms of a matrix model.  A great deal of technology has been
developed to deal with matrix models, primarily as a tool for studying
string theory.  The techniques we use here were derived in an earlier
pair of papers \cite{cot1,cot2}, and are related to methods described
in \cite{staudacher,dl}.  Some of the results which we describe here
appeared in a previous letter \cite{cot3}.

The Ising model on a regular lattice with a boundary magnetic field
was studied many years ago \cite{mw}. This model has been of renewed
interest recently because in the continuum limit, where the lattice
spacing is taken to zero, it gives a simple example of a
two-dimensional field theory which is conformal in the bulk but which
has boundary conditions breaking conformal invariance
\cite{gz,cz,klm}.  The only boundary conditions for the Ising model
which preserve conformal invariance \cite{cardy} are free boundary
conditions (where the boundary field $h$ vanishes), and fixed spin
boundary conditions (where $h = \pm \infty$).  Putting an arbitrary
field $h$ on the boundary generates a renormalization group (RG) flow
from the free boundary condition to the fixed boundary condition
\cite{al}.

For a matrix model corresponding to fields on a random surface, a
continuum limit can also be taken.  In this limit, the theory describes
conformal matter fields coupled to 2D quantum gravity.  In
this paper we take the continuum limit of the disk partition function
and magnetizations and consider the implications of our results for
the resulting theory of $c = 1/2$ matter coupled to gravity.  In particular, 
we find that the results are in accord with the hypothesis that the
RG flow which has been understood in flat space is present in an
appropriate form in the theory with gravity.

One provocative feature of our results is that as the boundary
magnetic field is increased, except for a jump discontinuity when the
field becomes nonzero, the expectation value of the magnetization of a
randomly chosen spin in the bulk {\it decreases}.  This
counterintuitive result may be explained in terms of the effects of
the matter fields on the geometry --- roughly speaking, the increase
in magnetic field produces a long ``throat'' which separates the
boundary from the bulk and which increases the average distance of a
bulk spin from the boundary, effectively decreasing the bulk
magnetization.  However, this result is also found to be a finite
volume effect which may depend upon the precise choice of how the
random lattice ensemble is defined.

Another context in which this work may be relevant is the current
discussion of D-branes in string theory (see for example
\cite{polchinski}).  Just as there are two conformally invariant
boundary conditions for the Ising theory, a conformal field theory of
a single bosonic field can have two conformally invariant boundary
conditions: Neumann and Dirichlet.  Considering the continuum limit of
the Ising model as a single free fermion, free and fixed Ising boundary
conditions are related through supersymmetry to Neumann and Dirichlet
boundary conditions on a bosonic field.  The fact that RG flow behaves
similarly in flat space and in the presence of a fluctuating metric
suggests that perhaps Dirichlet boundary conditions in superstring
theory naturally arise as an RG limit of a non-conformal boundary term.

In Section 2 of this paper we describe the discrete model we will use
for the Ising model on a random surface.  In Section 3 we apply the
methods of \cite{cot1} to this model, explicitly deriving
a quartic equation satisfied by the disk amplitude and locating the
associated critical point. 
In Section 4 we take the continuum limit of the disk
amplitude.  In Section 5 we discuss the general formalism we will use
for calculating magnetizations, and we apply this in Sections 6 and 7
to compute the boundary and bulk magnetizations on the disk.
In Section 8 we discuss implications of our results for renormalization
group flow, duality, and the effects of gravity on the behavior of
the matter theory.

\section{The model}
\label{sec:model}

A discretized theory of $c=1/2$ matter coupled to 2D quantum gravity
is described by the Ising model on a randomly triangulated surface.  At the
center of each (equilateral) triangle on the surface lives a single
Ising spin, coupled to its nearest neighbors.  This theory can be
described by a matrix model \cite{kazakov} with partition function
\begin{equation}
{\cal Z} (g, c)= \int {\rm D} U \; {\rm D} V \;
\exp \left(-NS (U,V) \right)\ ,
\label{eq:matrixising}
\end{equation}
where the action is given by 
\begin{equation}
  S={1\over{2}}\Tr\; (U^2+ V^2)-c \Tr\; UV
  -{g\over 3}\Tr\; (U^3+V^3)\ .
 \label{eq:origaction}
\end{equation}
In these expressions, $U$ and $V$ are $N\times N$ hermitian matrices
representing up and down spins respectively, $g$ is a coupling
constant corresponding to a Boltzmann weight for each triangle on the
surface, and $c$ describes the coupling between Ising spins.  (This
$c$ is unrelated to the central charge.)  The partition function can
be expanded in a power series in $g$ and $1/N$, and the coefficient of
$g^k N^{2-2h}$ is then given by a summation of the Ising partition
function over all triangulations of a genus $h$ Riemann surface by $k$
triangles.

We are interested in amplitudes corresponding to various boundary
conditions on spins living on a genus zero surface with boundary.  For
example, the disk amplitude for a configuration of plus and minus
spins on the boundary, represented by an ordered string $w(U,V)$ of
the matrices $U$ and $V$, is given by the large-$N$ limit of the
matrix model expectation value,
\begin{equation}
  p_{w} = \lim_{N\rightarrow \infty}{1\over N}\langle\Tr\; w(U,V)\rangle\ .
\end{equation}
(Since we will always be interested in the large-$N$ limit, henceforth
we will suppress the explicit $1/N$ in expectation values.)   Again,
$p_w$ can be expanded in a power series in $g$, with the coefficient
of $g^k$ giving the sum over all disk triangulations by $k$ triangles
with a boundary having a fixed length and spin configuration $w (+, -)$.

There are two types of conformally invariant boundary condition in the
Ising theory: ``fixed'' (corresponding to all boundary spins aligned)
and ``free'' (corresponding to an equally weight\-ed sum of all
possible boundary spin configurations).  For each of these we can
define a generating function which encodes the amplitudes.  Fixed
boundary conditions are described by
\begin{equation}
  \widetilde\phi_{\rm fixed}(u) = \sum_{k=0}^{\infty} 
  \langle\Tr\; U^k\rangle u^k\ ,
\end{equation}
while free boundary conditions are described by
\begin{equation}
  \widetilde\phi_{\rm free}(x) = \sum_{k=0}^{\infty} 
  \langle\Tr\; (U+V)^k\rangle x^k\ .
\end{equation}
The Ising model in this set of variables is symmetric under
interchange of $U$ and $V$ (corresponding to the symmetry under spin
reversal).  The generating function for fixed boundary conditions can
therefore also be described by summing over amplitudes with all $V$'s
on the boundary.

An alternative coupling of discretized $c=1/2$ matter to 2D quantum
gravity is obtained by considering the dual Ising model on a random
surface \cite{kostov}.  The partition function of the dual model is
defined once again as a sum over surfaces with Ising spins at the face
of each plaquette, where now the plaquettes may be polygons with
arbitrary numbers of sides, but the coordination number at each vertex
is constrained to be equal to three.  Such a configuration is
equivalent through duality to a triangulation in the original theory,
but with spins located at vertices rather than on faces.  The dual
model may also be described as a theory of two matrices $X$ and $Y$,
with action
\begin{equation}
  S =  \frac{(1-c)}{2} \Tr\; X^2 + \frac{(1 + c)}{2}
  \Tr\; Y^2 
  -\frac{\widehat{g}}{3} \Tr\;(X^3 + 3 XY^2)\ .
  \label{eq:dualaction}
\end{equation}
There are once again two types of conformally invariant boundary
conditions, fixed and free.  In the dual model the matrix variables
$X$ and $Y$ do not refer to the state of individual spins, but 
rather to the relative state of two spins across an edge; $X$ denotes
an edge separating two equal spins, while $Y$ denotes a boundary
between two opposite spins.  The generating function for fixed
boundary conditions is therefore
\begin{equation}
  \widehat\phi_{\rm fixed}(x) = \sum_{k=0}^{\infty} 
  \langle\Tr\; X^k\rangle x^k\ ,
\end{equation}
while free boundary conditions are described by
\begin{equation}
  \widehat\phi_{\rm free}(u) = \sum_{k=0}^{\infty} 
  \langle\Tr\; (X+Y)^k\rangle u^k\ .
\end{equation}
Note that the dual model is symmetric under $Y \rightarrow -Y$;
it is therefore the generating function for free boundary conditions
which has an alternative description in these variables, as a sum
over $\langle\Tr\; (X-Y)^k\rangle$.  (Amplitudes with an odd number of
boundary $Y$'s vanish identically.)

Although the original and dual formulations of the Ising model
represent different couplings to 2D gravity, there is a simple
transformation that relates the original action (\ref{eq:origaction}) 
to the dual action (\ref{eq:dualaction}):
\begin{eqnarray}
 X & \rightarrow &  \displaystyle \frac{1}{ \sqrt{2}}({U + V})
 \nonumber\\
 Y & \rightarrow &  \displaystyle \frac{1}{ \sqrt{2}}({U - V})
 \label{eq:transform}\\
 \widehat{g} &  \rightarrow &  \displaystyle g/{ \sqrt{2}}\ .\nonumber
\end{eqnarray}
As a consequence, the partition functions for the two models on
surfaces with no boundaries are identical.  
This does not, however, guarantee that all correlation functions in
the two theories agree,
since the transformation 
(\ref{eq:transform}) has a nontrivial action on the states and 
operators of the theory.  For example, the disorder operator $Y$ in the 
dual theory is taken into the spin operator $U-V$ in the
original theory.  Similarly, the role of free and fixed boundary
conditions is interchanged; we have
\begin{equation}\begin{array}{rcl}
  \widetilde\phi_{\rm fixed}(u) &=& \widehat\phi_{\rm free}(u/\sqrt{2})\ , \\
  \widetilde\phi_{\rm free}(x) &=& \widehat\phi_{\rm fixed}(\sqrt{2} x)\ .
  \end{array}
\end{equation}
The (Kramers-Wannier, or T-) duality of the model relates a specified
state (fixed, free, or with additional operator insertions) in the
original form of the model to the same state in the dual version.
Although in the discrete version of the theory, this duality symmetry
is explicitly violated by finite volume effects, it seems likely that
the duality symmetry is restored for all correlation functions in the
continuum limit.  Evidence for duality in the continuum limit at genus
zero was presented in \cite{sy,cot2}, and the issue of higher genus
was explored in \cite{asatani,kuroki,siegel}.

Our interest in this paper is in non-conformally-invariant boundary
conditions, which may be thought of as arising from the introduction
of boundary fields or couplings.  We therefore consider the original
model in a new set of matrix variables $Q$, $R$, defined by
\begin{equation}\begin{array}{rcl}
  Q &=& e^h U + e^{-h} V \ , \\
  R &=& e^h U - e^{-h}V \ .
  \end{array}
 \label{eq:qandr}
\end{equation}
Substituting these into the original action (\ref{eq:origaction}),
we obtain
\begin{equation}\begin{array}{rcl}
  S=&{1\over 4}\Tr\;\left[(\cosh(2h)-c) Q^2
  + (\cosh(2h)+c)R^2-2\sinh(2h) QR\right] \\
  &-{g\over 12}\Tr\; \left[\cosh(3h)(Q^3+3QR^2)-\sinh(3h)(3Q^2R+R^3)\right]\ .
  \end{array}
  \labell{eq:newaction}
\end{equation}
In this set of variables we can calculate the generating function
for disk amplitudes with all $Q$'s on the boundary,
\begin{equation}
  \phi(q,h) = \sum_{k=0}^{\infty} \langle\Tr\; (e^h U + e^{-h} V)^k\rangle q^k
  = \sum_{k=0}^{\infty} \langle\Tr\; Q^k\rangle q^k\ .
\label{eq:phi}
\end{equation}
This generating function describes boundary conditions which
interpolate between fixed and free.  The parameter $h$ can be thought
of as a boundary magnetic field applied to otherwise free boundary
conditions; for $h=0$ we recover $\widetilde\phi_{\rm free}$ in the
original model, while for $h=\pm\infty$ the boundary spins are all
driven to one value and we recover $\widetilde\phi_{\rm fixed}$. Note
that although the $h \to\pm \infty$ limit appears to be singular, this
is just an issue of normalization, which can be absorbed by a constant
rescaling of $q$. Indeed, we shall see that although the critical
value of $q$ goes to zero as $h \to\pm \infty$  with the
chosen normalization, all quantities of interest (such as $\phi$ for
example) are well behaved in these limits.

\section{Loop equations for correlation functions}
\label{sec:quartic}

The process of calculating the generating function $\phi(q)$ 
was described and essentially carried out in \cite{cot1},
without the explicit solution being written down.  Here we will
review the basics of that procedure, and examine the
solution in detail.  (The discussion in \cite{cot1} was framed in
terms of non-commuting variables; for our purposes here we may
skip directly to functions of a single variable.)

We begin by defining an additional set of generating functions
$\phi_{w(q,r)}(q)$.  These functions describe disks 
whose boundaries include a fixed string of matrices $w(Q,R)$,
plus any number of additional $Q$'s:
\begin{equation}
  \phi_{w(q,r)}(q) = \sum_{k=0}^{\infty} \langle\Tr\; w(Q,R)Q^k\rangle q^{k}\ .
 \label{eq:pwq}
\end{equation}
For example, we have
\begin{eqnarray}
  \phi_{rqr}(q)& =& \sum_{k=0}^{\infty} \langle\Tr\; RQRQ^k\rangle q^{k}
\\
&=& p_{rqr} + q p_{rqrq} + q^2 p_{rqrqq} + q^3
p_{rqrqqq} + \ldots \ .
\end{eqnarray}
 Recall that $p_{w(q,r)}$ is the amplitude for a disk
with boundary specified by the string $w(q,r)$; thus $p_q$  
corresponds to a single boundary edge labelled $q$, $p_r$ corresponds
to a single boundary edge labelled $r$, and $p_0= 1$ corresponds to
no boundaries.

We will also introduce a derivative operator $\D$ which acts on
power series in $q$.  Its effect is to annihilate terms independent of
$q$, and to remove one power of $q$ from all other terms:
\begin{equation}
  \D q^k = \cases{ 0 & for $k=0$\ ,\cr
  q^{k-1} & for $k\geq 1$\ .} \label{eq:D}
\end{equation}
The action of $\D$ on a generating function is to remove any constant
term and divide the remainder by $q$.  For example,
\begin{equation}
  \begin{array}{rcl}
  \D\phi &=& q^{-1}(\phi - 1) \ ,\\
  \D^2\phi &=& q^{-2}(\phi - 1-p_q q)\ , \\
  \D\phi_r &=& q^{-1}(\phi_r-p_r)\ , 
  \end{array}
  \label{eq:deriv}
\end{equation}
and so on. 

We can now derive a set of loop equations which relate these functions
to each other, by considering the possible outcomes of removing a
marked edge on the boundary.  For example, let us examine the effect
of removing the edge marked $R$ from the disks represented by
$\phi_r(q)$.  Since $\phi_r$ represents a sum over various
triangulated geometries of the disk, we can consider the effect of
removing $R$ from each of the terms separately.  For each term,
there are two basic possibilities: the edge might be
identified with another edge elsewhere on the boundary, or it might
belong to a triangle.  In the first case, removing the marked edge
and the one it was connected to leaves two disconnected triangulations,
both with all $Q$'s on the boundary and therefore representing $\phi(q)$.
In the second case, removing the triangle reveals two new boundary
edges, which may be marked with two $Q$'s, two $R$'s, or one $Q$ and
one $R$; these alternatives relate the initial generating function to
$\D^2\phi$, $\phi_{rr}$, and $\D\phi_r$.  This decomposition of
$\phi_r$ is shown schematically in Fig.~1.  The amount which each
term contributes to $\phi_r$ can be derived from the action
(\ref{eq:newaction}), as detailed in \cite{cot1}.

\begin{figure}
\centerline{
\psfig{figure=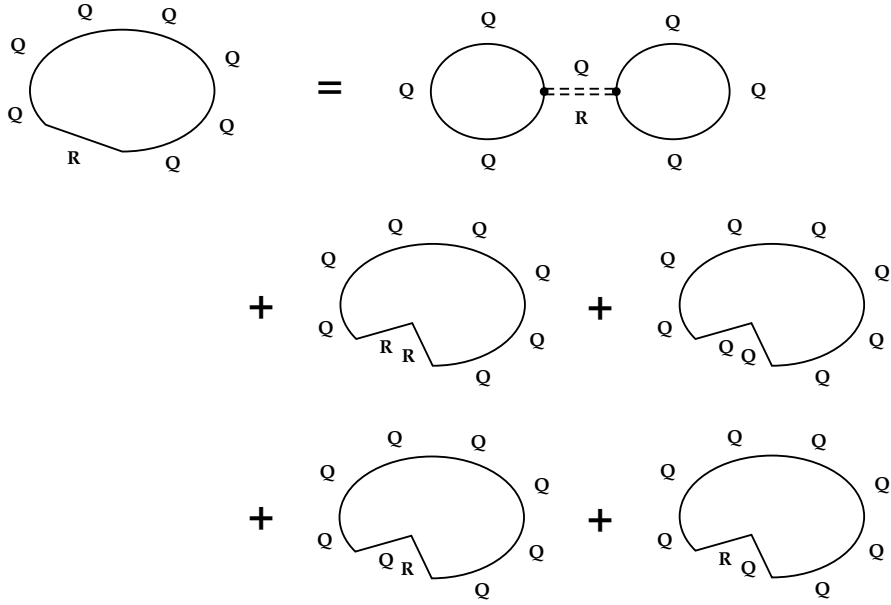,angle=0,height=8cm}}
\caption{Decomposition of $\phi_r$. Removal of the external edge
  marked $R$ on the left hand side leads to one of the possibilities
  shown on the right hand side, depending on whether that edge was
  connected to another exterior edge or an interior triangle.}
\end{figure}

By this procedure we can derive a closed system of eight 
independent equations in the 
quantities $(\phi,\phi_r,\phi_{rr},\phi_{rqr},\phi_{rrr},\phi_{rqqr},
\phi_{rqrr},\phi_{rrrr})$:
\begin{equation}
\begin{array}{rcl}
   \phi &=& 1+ \beta q^2 \phi^2 +a q \phi_{rr} +a \D\phi 
   + 2d \phi_r
    \\
 \phi_r &=& \gamma  q \phi^2+ b\phi_{rr} + b  \D^2\phi 
 + 2 e \D\phi_r
    \\
 \D\phi_r &=& \gamma  \phi + \beta q\phi\phi_r + 
   d \phi_{rqr} + a \phi_{rrr} + a \D^2\phi_r 
   + d \D\phi_{rr}
    \\
 \D^2\phi_r &=& \gamma  p_q \phi + \beta \phi_r 
   + \beta q\phi \D\phi_r+ 
   d  \phi_{rqqr} + a  \phi_{rqrr} + a  \D^3\phi_r + 
      d  \D\phi_{rqr}
    \\
 \D\phi_{rr}  &=& \gamma  p_r \phi + \gamma  \phi_r 
   + \beta q \phi \phi_{rr} + 
   d  \phi_{rqrr} + a  \phi_{rrrr} + a  \D^2\phi_{rr} + 
      d  \D\phi_{rrr}
    \\
 \phi_{rr} &=& \alpha\phi +\gamma q \phi\phi_r + e \phi_{rqr}
   + b\phi_{rrr} + b \D^2\phi_r +e\D\phi_{rr}
    \\
 \phi_{rqr} &=&  \alpha p_q \phi + \gamma  \phi_r + 
\gamma  q \phi \D\phi_r + e \phi_{rqqr} + 
   b \phi_{rqrr} + b  \D^3\phi_r + 
   e \D\phi_{rqr}
    \\
 \phi_{rrr} &=& \alpha p_r \phi + \alpha \phi_r + \gamma  
q \phi \phi_{rr} + 
   e \phi_{rqrr} + b \phi_{rrrr} + b  \D^2\phi_{rr} + 
   e \D\phi_{rrr}\ .
\end{array}
\labell{eq:gens}
\end{equation}

(In \cite{cot1} we listed ten equations in these variables, but the
last two were not linearly independent from the first eight.)
Here we have defined the new variables
\begin{equation}
  \begin{array}{rcl}
  \alpha &=& \frac{2}{1-c^2}[\cosh(2h)-c]\\
  \beta &=& \frac{2}{1-c^2}[\cosh(2h)+c]\\
  \gamma &=& \frac{2}{1-c^2}\sinh(2h)\\
  a &=& \frac{g}{2(1-c^2)} [\cosh{h} + c\cosh(3h)]\\
  b &=& -\frac{g}{2(1-c^2)} [\sinh{h} - c\sinh(3h)]\\
  d &=& -\frac{g}{2(1-c^2)} [\sinh{h} + c\sinh(3h)]\\
  e &=& \frac{g}{2(1-c^2)} [\cosh{h} - c\cosh(3h)]\ .
  \end{array}
\end{equation}
The set of equations (\ref{eq:gens}) is completely algebraic, since
we can replace the derivative operators with algebraic functions
as in (\ref{eq:deriv}).

Note that we can cut down somewhat on the number of undetermined
variables by looking at equations (\ref{eq:gens}) order by order.  These
give the following relations:
\begin{equation}\begin{array}{rcl}
  p_q &=& a p_{rr} + a p_{qq} + 2d  p_{qr}\\
  p_{qq} &=& \beta + a p_{qrr} + a p_{qqq} + 2d p_{qqr}\\
  p_r &=& b p_{rr} + b p_{qq} + 2ep_{qr}\\
  p_{qr} &=& \gamma  + b p_{qrr} + b p_{qqq} + 2e p_{qqr}\\
  p_{qr} &=& \gamma  + 2d p_{qrr} + a p_{rrr} + a p_{qqr}\\
  p_{rr} &=& \alpha + 2e p_{qrr} + b p_{rrr} + b p_{qqr}\ .
\end{array}
\label{eq:crels}
\end{equation}
Furthermore, each of these amplitudes may be expressed as a sum of
amplitudes in the original Ising model; for example, 
$p_q = e^hp_u + e^{-h}p_v = 2(\cosh{h}) p_u$.  We use these relations
to eliminate everything but $p_1=p_u$ and $p_3=p_{uuu}$.

It is now a straightforward but tedious exercise to solve the system
(\ref{eq:gens}) by deriving a single quartic equation for $\phi$ as a
function of $q$, $h$, $c$, $g$, $p_1$ and $p_3$.  The quartic is given
by
\begin{equation}
  f_4\phi^4+f_3\phi^3+f_2\phi^2+f_1\phi+f_0=0\ ,
  \label{eq:quartic}
\end{equation}
with
\begin{eqnarray}
&& \begin{array}{rcl}
 f_4 &=& 2g^2[1-\cosh(6h)] q^8\ ,
\end{array}  
\nonumber \\
&& \begin{array}{rcl}
 f_3 & = & -4 g^3 \cosh(3h)  q^5  \\
    &&+ 4g^2[\cosh(4h)+\cosh(2h)+c(3-\cosh(6h))]  q^6  \\
    &&+ 2g[c^2\cosh(9h) + c\cosh(7h) 
      - (1 + 3 c^2)\cosh(3h) - (1 + 5 c)\cosh(h)]  q^7\ ,
\end{array}
\nonumber   \\
&& \begin{array}{rcl}
 f_2 & = & -g^4 q^2 + 
     4g^3 [\cosh(h) - 2c \cosh(3h)]  q^3  \\
  & &+ g^2[4 c^2\cosh(6h) + 12 c\cosh(4h)
        - 2(1 - 5 c)\cosh(2h) - (3 - 23 c^2) ]  q^4  \\
  & &  + 2g[-2 c^2\cosh(7h) - 2c(1 + 2 c)\cosh(5h)
        - (5 c + 13 c^3)\cosh(3h)  \\
  & & \qquad  + (1 - 4 c - 19 c^2)\cosh(h)]  q^5  \\
  & &  + 2[c^3\cosh(8h) + c(2 c + 2 c^3 + g^2)\cosh(6h)\\
  & & \qquad + (c + 2 c^2 + 5 c^3 - g^2 + 2 p_1 g^3)\cosh(4h)  \\
  & & \qquad  + (2 c + 4 c^2 + 6 c^3 + g^2 - 2 p_1 g^3)\cosh(2h)
        + c(1 + 4 c + 2 c^3 - g^2) ]  q^6\ ,
\end{array}
\nonumber   \\
&& \begin{array}{rcl}
 f_1 & = & -2 c g^4   + 2c g^3 [5\cosh(h) + c\cosh(3h)] q  \\
  & &  - 2cg^2[2c\cosh(4h) + 4(1 + c)\cosh(2h)
        + (5 + 3 c^2) ]  q^2  \\
  & &  + 2g[c^2(1 + c)\cosh(5h) + c(1 + 6 c - 5 c^3 + 
        3 g^2)\cosh(3h)  \\
  & & \qquad  + (6 c + 2 c^2 - 4 c^3 - g^2 + 
        2 p_1 g^3)\cosh(h)]  q^3  \\
  & &  + 2[-c^2(c - c^3 + g^2)\cosh(6h)
        + 2c(-c + c^3 - 2 g^2 + p_1 g^3)\cosh(4h)  \\
  & & \qquad  + (-c - 2 c^2 + c^3 + 2 c^4 + g^2 - 2 p_1 g^3 
        - 6 c p_1 g^3)\cosh(2h)  \\
  & & \qquad  + (-c + c^5 + g^2 - 5 c^2 g^2 
        - 2 p_1 g^3) ]  q^4  \\ 
  & &  + 2g[c^2(1 - p_1 g)\cosh(7h) + c(1 - p_1 g + 
        2 c p_1 g)\cosh(5h)  \\
  & & \qquad  + 3c(c^2 + p_1 g)\cosh(3h) \\
  & & \qquad  + (-1 + 4 c^2 + 2 p_1 g + 2 c p_1 g + c^2 p_1 g)
            \cosh(h)]  q^5\ ,
\end{array}
\nonumber \\
&& \begin{array}{rcl}
 f_0 & = & 2 c g^4   + 2cg^3[-c\cosh(3h) +(- 5 + 2 p_1 g)
     \cosh(h)]  q  \\
  & &  + g^2[2c^2(2 - p_1 g)\cosh(4h)
     + 2c(4 + 4 c - 3 p_1 g - 3 c p_1 g)\cosh(2h)  \\
  & & \qquad  + (10 c - 6 c^3 + 4 p_1 g - 14 c p_1 g 
     - 3 g^2 - 4 p_3 g^3) ]  q^2  \\
  & &  + 2g[c^2(- 1 - c + p_1 g + c p_1 g)\cosh(5h)  \\
  & & \qquad  + c(-1 - 6 c + 5 c^3 - 3 p_1 g + 13 c p_1 g + 
     3 g^2 + 4 p_3 g^3)\cosh(3h)  \\
  & & \qquad  + (- 6 c - 2 c^2  + 4 c^3  - 
     4 p_1 g  + 13 c p_1 g  + 2 c^2 p_1 g \\
  & & \qquad \qquad + c^3 p_1 g
     + 3 g^2 +4p_3 g^3)\cosh(h)]  q^3  \\
  & &  + [2c^2(c- c^3 + p_1 g - 3 c p_1 g - g^2 - 
     p_3 g^3)\cosh(6h)  \\
  & & \qquad  + 2c(2 c - 2 c^3 + 2 p_1 g - 
     6 c p_1 g - g^2 - p_1 g^3 - 2 p_3 g^3)\cosh(4h)  \\
  & & \qquad  + 2(c + 2 c^2  - c^3  - 2 c^4  + p_1 g  - c p_1 g 
     - 6 c^2 p_1 g - g^2  - 2 c g^2  + p_1 g^3   \\
  & & \qquad\qquad  + c p_1 g^3  - 
     p_3 g^3  - 2 c p_3 g^3  - p_1^2 g^4)\cosh(2h)  \\
  & & \qquad  + (2 c - 2 c^5 + 2 p_1 g - 6 c p_1 g 
     + 2 c^2 p_1 g - 6 c^3 p_1 g - g^2 - c^2 g^2  \\
  & & \qquad\qquad  - 2 p_1 g^3 - 2 p_3 g^3 
     - 2 c^2 p_3 g^3 + 2 p_1^2 g^4) ]  q^4\ .
\end{array}
\nonumber \\
\end{eqnarray}

The analytic solutions to such an expression are of course rather
unwieldy, but fortunately they are also of little interest to us.
Instead, we are interested in the expansion of $\phi$ around the
critical point of the model, which encodes the continuum limit of
the theory.  (See \cite{cot2} for a discussion of the extraction
of the continuum limit.)  The critical values of the quantities
$c$, $g$, $p_1$ and $p_3$ are well known \cite{kazakov,cot2}:
\begin{equation}
  c_c=\frac{1}{ 27} {\left(-1 + 2 \sqrt{7} \right) }\ ,
  \label{ccrit}
\end{equation}
\begin{equation}
  g_c={3^{-9/2}\sqrt{10}\left(- 1 + 2 \sqrt{7} \right)^{3\over 2}}\ ,
  \label{gcrit}
\end{equation}
\begin{equation}
  g_c p_{1c}= \frac{1}{5}  ({3 -\sqrt{7}})\ ,
  \label{p1crit}
\end{equation}
and
\begin{equation}
  g_c p_{3c}=\frac{1}{100}( {-699 + 40\cdot 7^{3\over 2}})\ .
  \label{p3crit}
\end{equation}

We would now like to find the critical value of $q$ for
any given boundary field $h$.  The critical point is defined as
the radius of convergence of the power series expansion for the
generating function, interpreted physically as the point where
boundaries with an infinite number of segments begin to dominate.
The generating functions for fixed boundary conditions,
$\widetilde\phi_{\rm fixed}(u)$ in the original model and
$\widehat\phi_{\rm fixed}(x)$ in the dual model, obey cubic
equations, and the critical point is simply the value of $x$ or
$u$ for which the cubic has repeated roots.  For the quartic
this story is slightly more complicated.

Fig.~2 shows a numerically generated plot of the real part of the
solutions to the quartic, as a function of $q$, for $\exp(h)=5/3$ and
the other parameters at their critical values.  Only three distinct
curves are visible, as two of the solutions have identical real parts.
At three points on the graph the curves intersect, representing
multiple roots; as $q$ increases, there is first a triple root, then a
double root, and another triple root.  To decide which of these
represents the actual critical point, we follow the behavior of the
physical solution as $q$ is increased from zero along the real axis,
and look for a branch point (which will indicate the radius of
convergence).  {}From the definition of the generating function we
know that the physical solution is the one which equals one at $q=0$.
As $q$ is increased, the first triple root does not represent a branch
point for this solution, and is therefore not the critical point.  A
similar situation holds for the double root; even though the physical
solution is one of the double roots, one can verify numerically that
circling the double root in the complex $q$ plane does not take you to
a distinct Riemann sheet, and hence this value is not a branch point.
The critical value of $q$ is therefore at the second of the two triple
roots, as indicated by the point (c) in the figure.

\begin{figure}
\vskip -3.75cm
\centerline{
\psfig{figure=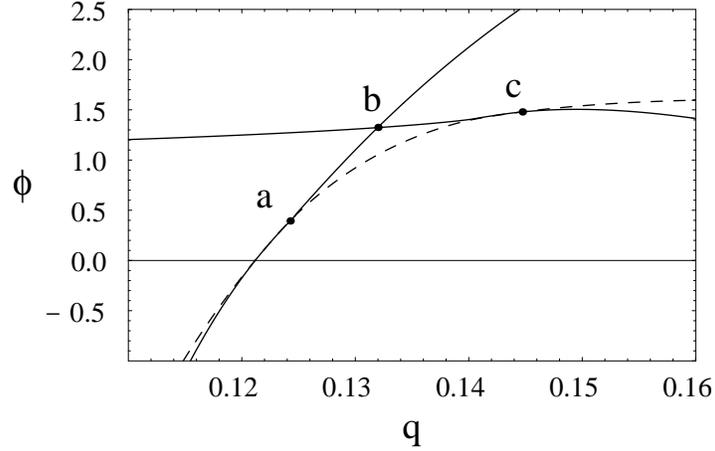,angle=0,height=13.5cm}}
\vskip -3cm
\caption{The four roots of the quartic, as a function of $q$.  The dashed line 
represents two roots which are complex conjugates of each other.  Moving
from left to right, we find the triple root ``a", the double root ``b",
and the triple root ``c" which is the critical point.  The physical
branch is the one that goes to 1 as $q$ goes to zero.}
\end{figure}

To discover an analytic expression for the critical value $q_c$, we note
that at this point the quartic can be factored into the form
\begin{equation}
  f_4(\phi - \phi^{(1)})(\phi - \phi^{(3)})^3 = 0\ ,
  \label{eq:qtriple}
\end{equation}
where $\phi^{(3)}$ is the triple root and $\phi^{(1)}$ is the single
root.  Setting the coefficients of ({\ref{eq:qtriple}}) equal to
those of (\ref{eq:quartic}) yields a set of equations from which we
can eliminate $\phi^{(1)}$  and $\phi^{(3)}$ to obtain
a single octic equation for $q_c$ as a function of
$h$.  Happily, this octic factors into the
product of two quadratics (one of which is cubed).  It is most conveniently
written in terms of $q_c/g_c$:
\begin{equation}\begin{array}{rcl}
  0 & = & [729 + (z_1 \cosh{3h} + z_2 \cosh{h})(q_c/g_c)
         + (z_3 \cosh{4h} + 2z_3 \cosh{2h} + z_4)(q_c/g_c)^2]^3 \\
    &  & \times [729 + (z_1 \cosh{3h} + z_2 \cosh{h})(q_c/g_c)
         + (z_5 \cosh{4h} + 2z_5 \cosh{2h} + z_6)(q_c/g_c)^2] \ ,
  \end{array}
  \label{eq:octic}
\end{equation}
where
\begin{equation}\begin{array}{rcl}
  z_1 &=& 54(1-2\sqrt{7}) \\
  z_2 &=& -1458\\
  z_3 &=& 6(10+7\sqrt{7}) \\
  z_4 &=& 2(262-11\sqrt{7}) \\
  z_5 &=& 18(-16+5\sqrt{7})\\
  z_6 &=& 2(784-83\sqrt{7}) \ .
  \end{array}
  \label{eq:ocvars}
\end{equation}
The critical point is the solution obtained by setting the
cubed quadratic to zero; one root of the quadratic will be the
critical point for $h\geq 0$, and the second for 
$h\leq 0$. For $h\geq 0$ the critical value is
\begin{equation}
  q_c(h) = \frac{g_c(1+2\sqrt{7})e^{3h}}{1+(
-1+\sqrt{7})e^{2h}
    +(2+\sqrt{7})e^{4h}}\ .
  \label{eq:critpts}
\end{equation}
whereas for $h\le 0$, 
\begin{equation}
  q_c(h) = \frac{g_c(1+2\sqrt{7})e^{-3h}}{1+(
-1+\sqrt{7})e^{-2h}
    +(2+\sqrt{7})e^{-4h}}\ .
\end{equation}
so that although $q_c(h)$ is continuous at $h=0$, it is non-analytic
there. 

The corresponding critical value of $\phi$ is
\begin{equation}
\phi_c(h) = {(3e^{2h}-1)(1 + (-1 + \sqrt{7})e^{2h} + 
(2+\sqrt{7})e^{4h})\over
        10\,e^{6h}}
\end{equation}
Note that as expected, $q_c\to 0$ as $h\to \infty$, but $\phi_c$ is
finite and non-zero for all $h$.

\section{Continuum limit and disk amplitude}
\label{sec:cont}

To extract information about the theory in the continuum limit, we
analyze the behavior of the discrete model in the vicinity of the
critical point.  The coupling constant $c$ is set to the critical
value $c_c$ given by (\ref{ccrit}); deviations from this value would
correspond to introducing a mass for the Majorana fermion in the
continuum limit of the Ising model, which we do not examine in this
paper.  We then trade the independent
variables $g$ and $q$ for new variables $t$ and $z$, defined by
\begin{equation}\begin{array}{rcl}
  g &=& g_c e^{-\epsilon^2 t}\ , \\
  q &=& q_c e^{-\epsilon z} \ 
  \end{array}
  \label{eq:gqexp}
\end{equation}
where $\epsilon$ is a small parameter indicating the distance from the
critical point.  The expansions of $p_1$ and $p_3$ in powers of
$\epsilon$ are then given by
\begin{eqnarray}
  gp_1&=&{{3 - \sqrt{7}}\over 5}
        \Biggl[ 1 - {{ 22 + 10 \sqrt{7}}\over
            9} \;  \epsilon^{2} t
  + {5^{1/3} (55 + 25 \sqrt{7} )
              \over 36}\;\epsilon^{8/3} t^{4/3}
  \nonumber\\
  && \quad + {5^{5/3} (11 + 5 \sqrt{7}) \over 216}\;
         \epsilon^{{10/3}} t^{5/3}\Biggr]  + {\cal O} (\epsilon^4t^{2})
  \label{P1exp}
\end{eqnarray}
and
\begin{eqnarray}
  gp_3&=& {-699 + 40 \cdot 7^{3/2}\over 100}  -
   {4 \left( 121 - 5 \cdot 7^{3/2} \right)\over 25} \;\epsilon^{2} t
  \nonumber\\
  && \quad+ {9 \cdot 5^{1/3} \over 2} \; \epsilon^{8/3} t^{4/3}
   + {3 \cdot 5^{2/3} \over 4}\; \epsilon^{{10/3}} t^{5/3}
  +{\cal O} (\epsilon^4 t^{2})\ .
  \label{P3exp}
\end{eqnarray}

These expansions may be substituted into the quartic (\ref{eq:quartic}),
which may then be solved for $\phi$ as a sum of increasing powers of
$\epsilon$.  We obtain
\begin{equation}
  \begin{array}{rcl}
  \phi &=& \displaystyle
  \phi_c(h)- \; {{ 3 + 2(-2 + \sqrt{7})e^{2h} + (1 + 
      2\sqrt{7})e^{4h}} \over 10e^{4h}}
  \;{\epsilon Z}
  \\
  &&\displaystyle
  \quad + \;
  {1 \over 5 \cdot 2^{7/3} \alpha(h)}
  \;{\epsilon^{4/3}\Phi} 
  + \; {2^{4/3}\,
        \left( 1 + 4e^{2h} + e^{4h} \right) \,
       \over 
    15\,\alpha(h)\,\left( 1 - e^{4h} \right)}
  \left[{\epsilon^{5/3}\,Z\,
      \left( 2T + Z^2 \right)\,\Phi \over {\Phi^2 - (4T)^{4/3}}}\right]
  + \; {\cal O}(\epsilon^2)\ .
  \end{array}
  \label{eq:phie}
\end{equation}
where for convenience we have rescaled the variables to
\begin{equation}
        T = 5t
\end{equation}
and
\begin{equation}
        Z = {z\over\alpha(h)}\ ,
\end{equation}
and the function $\Phi(Z,T)$ is given by
\begin{equation}
  \Phi(Z,T) \equiv \left(Z+\sqrt{Z^2-4T}\right)^{4/ 3}
  +\left(Z-\sqrt{Z^2-4T}\right)^{4/ 3}\ .
\end{equation}
For $h>0$, 
\begin{equation}
  \alpha(h)={e^{2h}\left(1+e^{2h}\right)\over 1 + (-1 + \sqrt{7})e^{2h} + 
  (2+\sqrt{7})e^{4h}}\ .
\end{equation}
but this function is discontinuous at $h=0$. If we take the continuum
limit in (\ref{eq:quartic}) after $h$ is set to zero (cf. the
calculation in \cite{cot2}), we find that $\phi$ is given by
(\ref{eq:phie}) with
\begin{equation}
\alpha(0) = {1\over \sqrt{2}(1+ \sqrt{7})}
\end{equation}

The universal part of $\phi$ is the first non-analytic term, and
appears at order $\epsilon^{4/3}$. By virtue of the discontinuity in
$\alpha(h)$, both the numerical coefficient of the universal term, and
the amplitude of $\phi$ as a function of $z$, are discontinuous.  

The universal part can be converted into the asymptotic form of the
disk amplitude $\tilde{\phi} (l, a)$ for fixed boundary length $l$ and
disk area $a$.  These forms of the amplitude are related through a
Laplace transform
\begin{equation}
  {1 \over 5 \cdot 2^{7/3} \alpha(h)}
  \;{\epsilon^{4/3}\Phi\left({z/\alpha(h)},5t\right)}
 =  \int {\rm d} l \int {\rm d} a \; \;e^{-zl-ta}
  \tilde{\phi} (l, a)\ .
\end{equation}
Inverting the Laplace transform, we have
\begin{equation}
  \tilde{\phi} (l, a) = {1\over 25\sqrt{3}\pi}
  (\alpha (h)l)^{1/3} (a/5)^{-7/3}
  e^{-5(\alpha (h)l)^2/a}  =
  {1\over 25\sqrt{3} \pi} L^{1/3} A^{-7/3} e^{-L^2/A}\ ,
\end{equation}
with the rescalings
\begin{equation}
  L = \alpha (h) l,\qquad A = {a\over 5}\ .
  \label{eq:rescale} 
\end{equation}
Up to an irrelevant multiplicative constant, this is precisely the
form of the disk amplitude when the boundary conditions are conformal
\cite{mss,staudacher,gn,cot2} (i.e., with $h = 0$ or $h = \pm
\infty$); however, the boundary length $l$ is rescaled by the factor $
\alpha (h)$ which depends discontinuously on the boundary magnetic
field.  Note that this amplitude includes an extra factor of $l$
corresponding to a marked point on the boundary.

\section{Expectation values}
\label{sec:exp}

Now that we have defined the continuum limit of the model and computed
the disk amplitude, we would like to calculate a number of correlation
functions related to the boundary and bulk magnetizations in the
theory.  In this section we describe the formalism necessary to
perform such calculations efficiently.

An example of the type of calculation we need to perform is the
limiting value of the boundary magnetization on a disk with a large
number of triangles and boundary segments.
The boundary magnetization for a spin on the boundary of a disk with
$k$ boundary edges and $n$ triangles is given by 
\begin{equation}
\langle m \rangle_{n,k} =
{\langle\Tr\;(e^h U - e^{-h}V)
(e^hU+e^{-h}V)^{k-1}\rangle_n\over\langle\Tr\;
(e^hU+e^{-h}V)^{k}\rangle_n}\ ,
\label{eq:bm}
\end{equation}
where by $\langle \rangle_n$ we indicate a sum over triangulations
restricted to geometries with $n$ spins (the coefficient of $g^n$ in
an expansion in $g$). The quantity $\langle m\rangle$ is defined to be
the large $n$ and $k$ limit of (\ref{eq:bm}).

More generally,
we define the expectation value of an operator in a specified
state $\psi$ as
\begin{equation}
  \langle {\cal A} \rangle = \lim_{k\to \infty,\; n\to \infty}
  {\langle {\cal A} \psi \rangle_{n,k} \over
  \langle \psi \rangle_{n,k}}\ ,
  \label{eq:expect}
\end{equation}
where $\langle \psi \rangle_{n,k}$ is taken to mean the sum over all
triangulations, with appropriate weights, with $n$ triangles and $k$
boundary edges. $\langle {\cal A}\psi \rangle_{n,k}$ is the same
quantity, but with the weights adjusted by the operator ${\cal A}$.

For the cases we are considering in this paper, $\psi$ is a sum over
all triangulations, with weights determined by the boundary
magnetic field. Thus, for zero field, $\psi$ is simply a
sum which is equally weighted for all
boundary configurations (free boundary conditions), whereas for
infinite (positive) $h$, the only boundary configurations with
non-zero weight are those with all spins pointing up (fixed boundary
conditions).  When ${\cal A}$ represents a boundary spin
operator, for example,
then for each configuration the boundary spin is evaluated
at a particular site, and the weight acquires a $\pm 1$ depending on
whether that spin is up or down. When ${\cal A}$ is a bulk spin
operator, the spin is evaluated at a site in the bulk.

The limits in (\ref{eq:expect}) can be understood in terms of the
asymptotic behavior of $\langle {\cal A} \psi \rangle_{n,k}$ and 
$\langle \psi \rangle_{n,k}$. 
For large $n, k$ these functions scale asymptotically as
\begin{equation}
  \langle \psi \rangle_{n,k} \sim g_c^{-n} q_c^{-k} f(n,k)
\end{equation}
and 
\begin{equation}
  \langle {\cal A} \psi \rangle_{n,k} \sim g_c^{-n} q_c^{-k} g(n,k)\ .
\end{equation}
Thus 
\begin{equation}
  \langle {\cal A} \rangle \sim
 \lim_{k\to \infty,\; n\to \infty}  {g(n, k)\over f(n, k)}\ .
\end{equation}
For large $n$ and $k$, it is appropriate to replace the number of
triangles and boundary edges by the area and length variables $a=
\epsilon^2 n$ and $l = \epsilon k$, so that $\langle{\cal A} \rangle$
will appear as a function of $a, l$.

The continuum limit of an operator expression of this type can be
easily determined by taking the continuum limits of the quantities
$\langle \psi \rangle$ and $\langle{\cal A} \psi \rangle$.
Consider the continuum limits of the sums 
\begin{equation}
  \sum_{n,k=0}^\infty\langle \psi \rangle _{n,k} g^n q^k 
\end{equation} 
and 
\begin{equation}
  \sum_{n,k=0}^\infty\langle {\cal A} \psi \rangle _{n,k} g^n q^k\ .
\end{equation} 
The universal behaviors of these two sums give the Laplace transforms
of the two functions $f(n,k)$ and $g(n,k)$ with respect to $z$ and
$t$. For example
\begin{equation}
\sum_{n,k=0}^\infty\langle \psi \rangle _{n,k} q^k g^n
\sim \int dk \int dn \; 
e^{-\epsilon^2 n t} e^{-\epsilon k z} f(n,k)
= F_u( t, z)\ ,
\end{equation} 
where the subscript ``$u$'' indicates the universal part.
In order to recover the functions $f$ and $g$, it suffices to perform
an inverse Laplace transform on the universal parts $F_u(
t, z)$ and $G_u( t, z)$ of the sums to obtain the 
functions
$\tilde{f}(a,l)$, for which
\begin{equation}
F_u( t, z) 
= \int d l \int d a \; e^{-z l - t a}
\tilde{f}(a,l)
\label{eq:tilde}
\end{equation}
and similarly $\tilde{g}(a,l)$. Thus
\begin{equation}
\langle {\cal A} \rangle = {\tilde{g}(a,l)\over \tilde{f}(a,l)}\ .
\end{equation}
We shall see this explicitly in the following sections.

\section{Boundary magnetization}
\label{sec:bomf}

In this section we will apply the discussion above to compute
the one- and two-point boundary magnetizations in the presence 
of a boundary field.

\subsection{One-point boundary magnetization}

The boundary magnetization $\langle m \rangle$ is given by  the large
$n, k$ limit of
\begin{equation}
\langle m \rangle_{n,k} =
{\langle\Tr\;(e^h U - e^{-h}V)
(e^hU+e^{-h}V)^{k -1}\rangle_n\over\langle\Tr\;
(e^hU+e^{-h}V)^{k}\rangle_n}\ .
\label{eq:bm2}
\end{equation}

We may follow the route described in the previous
section to compute $\langle m\rangle$. The first step, the critical
expansion of $\phi$ in the continuum limit, was given in
(\ref{eq:phie}). The other quantity we need to expand is
\begin{equation}
  \psi_r \equiv 
  q\phi_r=\sum_{k =0}^\infty \langle\Tr\; RQ^{k}\rangle q^{k+1} \ .
\end{equation}
When $h = 0$, $\phi_r$ vanishes by symmetry.  When $h \neq 0$, we can
compute $\phi_r(h)$ by solving a linear
combination of the first two loop equations of (\ref{eq:gens}).
{}From these it follows that
\begin{equation}
  \phi_r = {{\left( 1 - e^{2h} \right) 
      \left[ (c - e^{2h} + ce^{2h} + ce^{4h})(1-\phi) + 
        2p_1{e^{2h}}g   - 2\phi^2q^2(1+e^{2h}+e^{4h}) \right] }\over 
    {cq + e^{2h}q + e^{4h}q + 
      ce^{6h}q -2e^{3h}g}}\ ,
\end{equation}
and hence, expanding and multiplying by $q_c e^{-\epsilon z}$,  
we obtain
\begin{eqnarray}
  \psi_r
  &=&       {{\left({e^{2h}-1} \right) 
       \left( -1 + (3 - \sqrt{7})e^{2h} - (4 - 
         3\sqrt{7})e^{4h} \right) }\over {10 \, {e^{6h}}}} 
  \nonumber \\
  && - {{\left({e^{2h}-1} \right) 
       \left( 3 + (1 + 2{\sqrt{7}}){e^{2h}} \right) }\over 
     {10 \, {e^{6h}} }} \; {\epsilon\, Z}
  \nonumber\\
  && + 
   {{\left(e^{2h}-1 \right) 
       \left( 3 + (2 + \sqrt{7})e^{2h} \right)}
      \over {5\cdot 2^{7/3}e^{2h}\left( 1 + {e^{2h}} \right) }
     } \; \epsilon^{4/3}\, \Phi 
  \label{eq:Phir}
  \\
  && -
   {{  2^{4/3}\left( 3 + (2 + \sqrt{7})e^{2h} \right) 
       \left( 1 + 4{e^{2h}} + {e^{4h}} \right) }\over 
     {15{e^{2h}}{{\left( 1 + {e^{2h}} \right) }^2}
       }}\left[{{\epsilon^{5/3} \, Z\,
       \left( 2T + {Z^2} \right)\, \Phi} \over {\Phi^2 -
         (4T)^{4/3}}} \right]
  \nonumber\\
  && + {\cal O}(\epsilon^2)\ .\nonumber
\end{eqnarray}
The universal part of $\psi_r$ is equal to that of $\phi$, up to an
$h-$dependent constant.  There is therefore no need to explicitly 
compute $\tilde\psi_r(a,l)$, the inverse Laplace transform of 
$\psi_r(Z,T)$, in order to determine the boundary
magnetization. We need only compute the ratio of the universal parts
to get the $l$ and $A$ independent result
(for $h > 0$)
\begin{equation}
  \langle m \rangle={\tilde\psi_r\over\tilde{\phi}}=
  {\left(e^{2h} - 1\right)\left(3 + (2+\sqrt{7})e^{2h}\right)\over
  \left(1 +(-1+\sqrt{7})e^{2h} + (2+\sqrt{7})e^{4h}\right)}\ .
  \label{eq:boundarymagnetization}
\end{equation}
Note that $\langle m \rangle$ is independent of $l$ and $A$, and is
continuous at $h=0$. (Note also that the expression given in
\cite{cot3} contains a typographical error in the numerator.)

\begin{figure}
\vskip -3.75cm
\centerline{
\psfig{figure=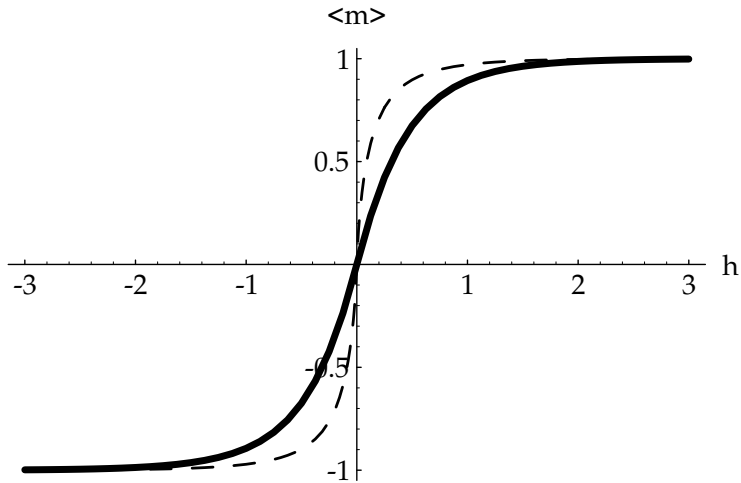,angle=0,height=13.5cm}}
\vskip -3cm
\caption{Boundary magnetization $\langle m \rangle$ as a function of
boundary field $h$ in flat space (dotted line) and on a random surface
(bold line)}
\end{figure}

In this particular case, there happens to be a simple argument that
gives $\langle m\rangle$ more directly than the computation outlined
above. In the large $k$ limit
\begin{equation}
\langle Q^k\rangle_{n}\sim q_c(h)^{-k} g_c^{-n} f(n,k)\ .
\end{equation} 
Differentiating both sides with respect to $h$, we obtain
\begin{equation}
k\langle Q^{k-1} R\rangle_{n}\sim k \left[ -
q_c(h)^{-(k+1)}q_c'(h) f(n,k)
+ {\cal O}({1/k})\right]g_c^{-n}\ ,
\end{equation}
from which it follows directly that in the large $k$ limit, 
\begin{equation}
\langle m\rangle =  -{q_c'(h)\over q_c(h)} =
{{\left(e^{2h} -1\right) 
      \left( 3 + 2e^{2h} + \sqrt{7}e^{2h} \right) }\over 
    1 + (-1 + {\sqrt{7}})e^{2h} + (2 + 
      {\sqrt{7}})e^{4h}}\ ,
\label{eq:mshort}
\end{equation}
where we have used (\ref{eq:critpts}) for $q_c(h)$. 
This confirms the result obtained in (\ref{eq:boundarymagnetization}).

A graph of the boundary magnetization is shown in Fig.~1 (bold
curve).  As expected, with no field the magnetization is zero,
and for an infinite field the magnetization is 1.  This
result is compared with the boundary magnetization on a half-plane
in flat space, computed by McCoy and Wu \cite{mw} (dashed curve).  
Whereas in flat
space the magnetization scales as $h \ln h$  for small $h$, leading to
a divergence in the magnetic susceptibility at the critical
temperature,  on a random surface we find that the magnetization is
linear at $h = 0$, with a finite susceptibility
\begin{equation}
\chi =\partial_h \langle m \rangle |_{h = 0} = \frac{1 + 2
\sqrt{7}}{3} \ .
\label{eq:ss}
\end{equation}

\subsection{Two-point boundary magnetization}

Having computed the magnetization at a single
point on the boundary of the disk in the presence of a boundary
magnetic field, we would now like to compute the correlation between
two spins on the boundary of the disk, which are separated by $k$ and
$l$ edges in the two directions around the boundary.
To compute the two point magnetization, 
\begin{equation}
  \langle m^2 \rangle =
  {\langle\Tr\;(e^h U - e^{-h}V)
  (e^hU+e^{-h}V)^{k}(e^h U - e^{-h}V)
  (e^hU+e^{-h}V)^{l}\rangle_n\over\langle\Tr\;
  (e^hU+e^{-h}V)^{k+l+2}\rangle_n}\ ,
\end{equation}
we need expressions for
\begin{equation}
  \Sigma(q_1,q_2)\equiv q_1 q_2\sigma(q_1,q_2)
\end{equation}
where
\begin{equation}
  \sigma(q_1,q_2)=
  \sum_{k =0}^\infty \sum_{l =0}^\infty
  \langle\Tr\; RQ^kRQ^l\rangle q_1^{k} q_2^{l}\ ,
\end{equation}
and for
\begin{equation}
  \rho(q_1,q_2)=\sum_{k =0}^\infty \sum_{l =0}^\infty
  \langle\Tr\; Q^{k+l+2}\rangle q_1^{k+1} q_2^{l+1}\ .
\end{equation}
Again, using the loop equation
techniques of \cite{cot1}, one can derive the
following set of equations for $\sigma$ (here, a
derivative $D_{q_i}$ denotes a combinatorial derivative as in
(\ref{eq:D}), where all variables other than $q_i$ are held constant):
\begin{eqnarray*}
\sigma(q_1,q_2) &=&{ b}{ \DDD^2\bar{\phi}(q_1,q_2)} + 
    e \left( { \DDD\sigma(q_1,q_2)} + 
        { \DD\sigma(q_1,q_2)} \right)  + 
    \alpha { \phi(q_1)}{ \phi(q_2)} 
\\
&&+ 
    \gamma { \bar{\phi}(q_1,q_2)}\left( { \phi(q_1)}{ q_1} + 
       { \phi(q_2)}{ q_2} \right)  + { b}{\sigma_r(q_1,q_2)}
\\
{ \bar{\phi}(q_1,q_2)} &=&
   { b}{ \DD^2\phi(q_2)} + 2 e{ \DD\phi_r(q_2)} + 
    { b}{\phi_{rr}(q_2)} + { a}{ \DDD^2\bar{\phi}(q_1,q_2)}{
q_1} 
\\
&&+ 
    { d}\left( { \DDD\sigma(q_1,q_2)} + 
        { \DD\sigma(q_1,q_2)} \right) {
q_1} + \gamma { \phi(q_1)}{ \phi(q_2)}{ q_1} 
\\
&&+ 
    \gamma {{{ \phi(q_2)}}^2}{ q_2} + 
    \beta { \bar{\phi}(q_1,q_2)}{ q_1}
     \left( { \phi(q_1)}{ q_1} + { \phi(q_2)}{ q_2} \right)  + 
    {a}{ q_1}{\sigma_r(q_1,q_2)}
\\
{ \phi(q_1)} &=&
1 + { a}{ \DDD^2\phi(q_1)}{ q_1} + 
    2{ d}{ \DDD\phi_r(q_1)}{ q_1} + {a}{\phi_{rr}(q_1)}{ q_1} + 
    \beta {{{ \phi(q_1)}}^2}{{{ q_1}}^2}   
\\
{ \phi(q_2)} &=&
1 + { a}{ \DD^2\phi(q_2)}{ q_2} + 
    2{ d}{ \DD\phi_r(q_2)}{ q_2} + {a}{\phi_{rr}(q_2)}{ q_2} + 
    \beta {{{ \phi(q_2)}}^2}{{{ q_2}}^2}
\\
{\phi_r(q_1)} &=&
   { b}{ \DDD^2\phi(q_1)} + 2 e \DDD\phi_r(q_1) + 
    { b}{\phi_{rr}(q_1)} + \gamma {{{ \phi(q_1)}}^2}{ q_1}
\\ 
{\phi_r(q_2)} &=&
{ b}{ \DD^2 \phi(q_2)} + 2 e{ \DD \phi_r(q_2)} + 
    { b}{\phi_{rr}(q_2)} + \gamma {{{ \phi(q_2)}}^2}
{ q_2} \ ,
\end{eqnarray*}
where 
\begin{eqnarray}
  \sigma_r(q_1,q_2)&=&\sum_{k =0}^\infty \sum_{l =0}^\infty
  \langle\Tr\; RQ^kRQ^lR\rangle q_1^{k} q_2^{l}
\end{eqnarray}
and
\begin{equation}
  \bar{\phi}(q_1,q_2)=\sum_{k =0}^\infty \sum_{l =0}^\infty
  \langle\Tr\; Q^{k}RQ^l\rangle q_1^{k} q_2^{l}\ .
\end{equation}
It is easily verified  that $\bar{\phi} (q_1, q_2)$
is given by
\begin{equation}
  \bar{\phi}(q_1,q_2)={q_1\DDD\phi_r(q_1) 
  - q_2\DD\phi_r(q_2)\over q_1 - q_2}\ .
\end{equation}
An expression for $\sigma(q_1,q_2)$ in terms of $\phi(q_1)$,
$\phi(q_2)$ and $p_1$ and $p_3$ can be obtained by solving these
equations. The expression is too long to be included here. 
On the other hand
$\rho(q_1,q_2)$ can be directly expressed in terms of $\phi(q_1)$ and
$\phi(q_2)$ in a very simple way as
\begin{equation}
\rho(q_1,q_2)={q_2 q_1^2 \DDD^2\phi(q_1)-q_1 q_2^2
\DD^2\phi(q_2)\over q_1-q_2} = 1+
{q_2\phi(q_1)-q_1\phi(q_2)\over q_1-q_2}\ .
\end{equation}

Armed with expressions for $\sigma(q_1,q_2)$ and $\rho(q_1,q_2)$, it
is then straightforward, if rather tedious, to obtain the critical
expansions of $\Sigma$ and $\rho$. They are given by
\begin{equation}
  \Sigma(Z_1,Z_2,T) =\sigma_c + {\left(e^{2h}-1\right)^2
  \left(3+(2+\sqrt{7})e^{2h}\right)^2\over
  20\cdot 2^{1/3} e^{4h} \left(1+e^{2h}\right)^2}
  \left[{{\Phi(Z_1,T)-\Phi(Z_2,T)}\over
  {Z_1 - Z_2}}\right] \epsilon^{1/3}
  + {\cal O}(\epsilon^{2/3})
\end{equation}
and
\begin{equation}
  \rho(Z_1,Z_2,T) = \rho_c +
  {1\over 20 \cdot 2^{1/3}\, \alpha(h)^2}
  \left[{{\Phi(Z_1,T)-\Phi(Z_2,T)}\over
  {Z_1 - Z_2}}\right] \epsilon^{1/3}
  + {\cal O}(\epsilon^{2/3})\ .
\end{equation}

As in the case of the one-point magnetization, the universal parts
of $\Sigma$ and $\rho$ depend on $Z_1$, $Z_2$, and $T$ in the same
way.  Consequently, the ratio of the Laplace transforms of these
universal parts will simply be the $h$-dependent ratio of the 
universal parts themselves.
It follows that the two-point boundary magnetization,
\begin{equation}
\langle m^2 \rangle =
{\tilde\Sigma\over\tilde\rho}=
{\left( e^{2h} - 1  \right)^2
       \left( 3 + (2+\sqrt{7})e^{2h}\right)^2
       \over \left(1 + (-1 + \sqrt{7})e^{2h} + 
        (2+\sqrt{7})e^{4h}\right)^2}\ ,
\end{equation}
is precisely the square of the one-point magnetization.  We find an
absence of polynomial corrections to this result (at least to the
first subleading order), indicating that correlations between boundary
spin operators decay exponentially, as one would expect in analogy
with the flat space theory.

\section{Bulk magnetization}
\label{sec:bmf}

The expectation value of the bulk magnetization in
the presence of a boundary magnetic
field, on a disk with boundary length $k$ and area $n$,
is given by
\begin{equation}
  \langle M\rangle =
  {\langle\Tr\;
  (e^hU+e^{-h}V)^{k}\cdot \Tr\;(U-V)\rangle_n\over\langle\Tr\;
  (e^hU+e^{-h}V)^{k}\cdot \Tr\;(U+V)\rangle_n}\ .
 \label{eq:bulk}
\end{equation}
This can be evaluated by considering cylinder amplitudes with one
boundary having a boundary magnetic field, and the other with a single
boundary edge.  The second boundary represents a marked point on the
bulk.  Although the second boundary corresponds to  only a single edge
rather than 3 edges as would be appropriate for a triangle
corresponding to a single spin, this distinction should not be
relevant in the continuum limit where the boundary becomes pointlike.
Again, a quantity such as (\ref{eq:bulk}) can be computed by the method of loop
equations \cite{cot2}.

To compute the magnetization, we require two punctured-disk amplitudes:
\begin{equation}
  \tau(h)=\sum_{k =0}^\infty \langle{\rm Tr}\;
  (e^hU+e^{-h}V)^{k}\cdot {\rm Tr}\;(U-V)\rangle q^k 
\end{equation}
and
\begin{equation}
  \lambda(h)=\sum_{k =0}^\infty \langle{\rm Tr}\;
  (e^hU+e^{-h}V)^{k}\cdot {\rm Tr}\;(U+V)\rangle q^k \ .
\end{equation}
As in (\ref{eq:pwq}), we define functions related to $\tau$ but with
additional words corresponding to sequences of spins on the outer
boundary:
\begin{equation}
  \tau_{w(q,r)}(q) = 
  \sum_{k =0}^\infty \langle{\rm Tr}\; w (Q, R)
  Q^{k}\cdot {\rm Tr}\;(U-V)\rangle q^k \ .
 \label{eq:twq}
\end{equation}

The first step in computing the bulk magnetization is now to derive a
set of eight independent equations which close on the quantities
$(\tau,\tau_r,\tau_{rr},\tau_{rqr},\tau_{rrr},\tau_{rqqr},
\tau_{rqrr},\tau_{rrrr})$:
\begin{equation}
\begin{array}{rcl}
\tau &=& 2 \beta q^2 \phi\tau +
 a q \tau_{rr} +a \D\tau 
   + 2d \tau_r + c_q q \phi
    \\
\tau_r &=& 2\gamma  q \phi\tau+ b\tau_{rr} + b  \D^2\tau 
 + 2 e \D\tau_r + c_r \phi
    \\
 \tau_{rr} &=& \alpha\tau +\gamma q \left(
        \tau\phi_r + \phi\tau_r\right) + e \tau_{rqr}
   + b\tau_{rrr} + b \D^2\tau_r +e\D\tau_{rr} + c_r \phi_r
    \\
\D\tau_r &=& \gamma  \tau + 
        \beta q\left(\tau \phi_r + \phi \tau_r\right)
   + d \tau_{rqr} + a \tau_{rrr} + a \D^2\tau_r 
   + d \D\tau_{rr} + c_q \phi_r
    \\
\D^2\tau_r &=& \gamma  \left(p_q \tau + t_q\phi\right)
        + \beta \tau_r 
   + \beta q\left(\tau \phi_r + \phi\tau_r\right) +
\\
&&
\qquad
   d \tau_{rqqr} + a  \tau_{rqrr} + a  \D^3\tau_r + 
      d  \D\tau_{rqr} + c_q \D \phi_r
    \\
 \D\tau_{rr}  &=& \gamma  \left(p_r \tau + t_r\phi\right)
+ \gamma  \tau_r 
   + \beta q \left(\tau \phi_{rr} + \phi \tau_{rr}\right) +
\\
&&
\qquad
   d  \tau_{rqrr} + a  \tau_{rrrr} + a  \D^2\tau_{rr} + 
      d  \D\tau_{rrr} + c_q \phi_{rr}
    \\
\tau_{rqr} &=&  \alpha \left( p_q \tau + t_q \phi\right)
   + \gamma  \tau_r + \gamma  q \left(\tau \D\phi_r + 
        \phi \D\tau_r \right) +
\\
&&
\qquad
        e \tau_{rqqr} + 
   b \tau_{rqrr} + b  \D^3\tau_r + 
   e \D\tau_{rqr} + c_r\D\phi_r    
\\
 \tau_{rrr} &=& \alpha \left(p_r\tau + t_r\phi\right)
+ \alpha \tau_r + \gamma  
q \left(\tau \phi_{rr} + \phi\tau_{rr}\right) +
\\
&&
\qquad
   e \tau_{rqrr} + b \tau_{rrrr} + b  \D^2\tau_{rr} + 
   e \D\tau_{rrr} + c_r\phi_{rr}\ ,
\end{array}
\labell{eq:genstau}
\end{equation}
where 
\begin{equation}
t_q\equiv\langle\Tr\; Q \cdot \Tr\;(U-V)\rangle,\qquad 
t_r \equiv \langle\Tr\; R \cdot \Tr\;(U-V)\rangle\ .
\end{equation}
As before, $p_q=\langle \Tr\; Q \rangle$, $p_r=\langle \Tr\;
R\rangle$, and the various constants take the same values as in
(\ref{eq:gens}).

The equations (\ref{eq:genstau}) can now be used to express $\tau(h)$ 
as a polynomial function of $\phi(h)$. However, this equation will also
contain a number of unknown correlation functions $t_q$, $t_{qq}$,
$t_{qqq}$, $t_r$, $t_{rr}$ and  $t_{rrr}$, some of which appear explicitly 
in (\ref{eq:genstau}) and some of which arise from the derivatives of 
$\tau$, since
\begin{equation}
  \begin{array}{rcl}
  \D\tau &=& q^{-1}(\tau - 1) \ ,\\
  \D^2\tau &=& q^{-2}(\tau - 1-t_q q)\ , \\
  \end{array}
  \label{eq:derivtau}
\end{equation}
and so on. As in the computation of $\phi$, these correlation
functions can be reduced to a much smaller number of unknowns by 
expanding (\ref{eq:genstau}) order-by-order. It turns out that 
after using all the relations in
(\ref{eq:genstau}) ({\it cf.} (\ref{eq:crels})), two extra relations
are required between
\begin{equation}
\begin{array}{l}
t_u\equiv\langle\Tr\; U \cdot \Tr\; U\rangle, \qquad 
t_{uuu}\equiv\langle\Tr\; U^3 \cdot \Tr\; U\rangle,  
\\
t_v\equiv\langle\Tr\; V \cdot \Tr\; U\rangle, \qquad
t_{vvv}\equiv\langle\Tr\; V^3 \cdot \Tr\; U\rangle.
\end{array}
\end{equation}
The first extra relation comes from the calculation in \cite{cot2}
of the critical expansion of
\begin{equation}
w_0\equiv {1\over 2}\langle \Tr\; (U-V) \cdot \Tr\; (U-V)\rangle
=
t_u - t_v\ ,
\end{equation}
which is given by
\begin{equation}
w_0= {1+2 \sqrt{7}\over 5} 
\left(1+5^{2/3}  \epsilon^{4/3} t^2 + {\cal O}(\epsilon^2 t^3)\right)\ .
\end{equation}
The second relation is obtained by differentiating the matrix integral 
expression for $p_1$ with respect to $g$:
\begin{equation}
\partial_g p_1  ={t_{uuu} + t_{vvv}\over 3}\ .
\end{equation}
Then
\begin{equation}
\partial_g p_1 = {\partial_tp_1\over -3 g \epsilon^2 t^2}
\end{equation}
can be used to obtain the critical expansion of $\partial_g p_1$.

Armed with these extra relations, we have all the information required
to expand $\tau$ in $\epsilon$. We obtain
\begin{equation}
\tau = {2^{4/3} g_c (1 + 2\sqrt{7})^2
        (1 +(-1+\sqrt{7}) e^{2h} + (2+\sqrt{7})e^{4h}) Z\over
        15 e^{2h}(1 + e^{2h}) (\Phi(Z,T) + (4T)^{2/3})} \epsilon^{-1/3} 
  + {\cal O} (\epsilon^{0})\ .
\end{equation}
As discussed in Sec.~\ref{sec:exp}, the quantity of interest
is the inverse Laplace transform of the universal part of $\tau$,
given in this case by 
\begin{equation}
\tilde{\tau}(L,A)=  
{(1 + 2\sqrt{7})^2 g_c  \over
50\sqrt{3}\pi} L^{2/3} A^{-5/3} e^{-L^2/A} \ .
\end{equation}
Here, we have introduced rescaled area and boundary length parameters
as in (\ref{eq:rescale}).

It is much easier to compute the critical expansion of $\lambda(h)$
since it is directly related to $\phi(h)$ via
\begin{equation}
  \lambda(h) = \partial_g\phi(h)\ .
\end{equation}
This gives an expansion
\begin{equation}
  \lambda =
  {2^{2/3} (-1 + 2\sqrt{7})(1 +(-1+\sqrt{7}) e^{2h} + (2+\sqrt{7})e^{4h})
  \Lambda(Z,T)
  \over 81 g_c e^{2h} (1+e^{2h})} \epsilon^{-2/3}
  + {\cal O}(\epsilon^{-1/3})\ ,
\end{equation}
where
\begin{equation}
  \Lambda(Z,T) = {
  (Z - \sqrt{Z^2 -4T})^{1/3} - (Z + \sqrt{Z^2 -4T})^{1/3}\over
  \sqrt{Z^2 -4T}}\ .
\end{equation}
The leading term in $\lambda$ has an inverse Laplace transform
\begin{equation}
\tilde{\lambda}(L,A) =
{(-1 + 2\sqrt{7})\over 5\cdot 3^{7/2}\pi 
g_c}L^{1/3} A^{-4/3} e^{-L^2/A}\ .
\end{equation}

The bulk magnetization in the continuum limit is then given by
\begin{equation}
  \langle M \rangle={\tilde{\tau}(L,A)\over\tilde{\lambda}(L,A)} =
  L^{1/3}A^{-1/3}\ .
  \label{eq:bulkmagnetization}
\end{equation}
The numerical values of the coefficients of $\tilde{\tau}$
and $\tilde{\lambda}$ have exactly cancelled.  Although it may appear
that the magnetization can be greater than one, this formula
is valid in the continuum limit, for which $A\sim L^2 \gg 1$.  We also
notice that this form of the magnetization is independent of $h$ except
for the dependence on the scaling factor $\alpha(h)$ incorporated in
$L$.  At $h = 0$, this magnetization is discontinuous and vanishes.
One nice feature of this result is that it correctly reproduces the
scaling behavior expected after the magnetization operator has been
gravitationally dressed according to the KPZ/DDK description of
Liouville theory \cite{kpz,ddk}.  The gravitationally dressed scaling
dimension of the bulk magnetization field is $\Delta = 1/6$. By
analogy with the flat space theory we expect that the bulk
magnetization should scale as $\langle M \rangle  \sim d^{-2 \Delta}$
where $d$ is a measure of the distance from the boundary.  This is
precisely the behavior seen in (\ref{eq:bulkmagnetization}), since
$A/L$ has dimensions of length.

\section{Discussion}

Let us summarize the implications of the results we have derived
in the previous sections.

\subsection{Renormalization group flow}

In Section 4 we computed the disk amplitude in the presence of a
boundary magnetic field as a function of the disk area $a$, the
boundary length $l$, and the boundary field $h$.  We discovered that
the result could be written in terms of the two variables $A=a/5$ and
$L=\alpha(h)l$, in which case the disk amplitude took on precisely the
form of the analogous function when the boundary conditions are
conformally invariant.  Thus, the effect of a boundary field on this
amplitude amounts to a rescaling of the boundary length by an
$h$-dependent factor.  In Section 7, meanwhile, we found that the bulk
magnetization as a function of $a$, $l$ and $h$ could also be
expressed in terms of $A$ and $L$, and that its functional form
was the same as in the presence of an infinite boundary field.
 
Taken together, these results imply the existence of an RG flow to the
conformally invariant boundary condition with an infinite boundary
field; $h$ is a relevant operator which goes to $\pm\infty$ in the
infrared (in this context, as the disk area and length grow large).
Further evidence is provided by the discontinuity in the rescaling
function at $h=0$:  any imposed boundary field, no matter how
small, leads to magnetization in the bulk of the same form as
that expected in the presence of fixed boundary conditions.

A related phenomenon has previously been derived for the flat-space
Ising model on a half-plane geometry \cite{mw,gz,cz}.  Again, one
can compute the magnetization of a point in the bulk in the presence
of a boundary field; however, rather than depending on the area of
the surface and length of the boundary (both of which are infinite
for the half-plane), the magnetization is a function of the distance
from the boundary.  (In quantum gravity, where we sum over all geometries, 
it would be conceivable but much more difficult to compute any quantity
as a function of, say, minimum geodesic distance from the boundary.
Computing anything ``at a fixed point'' is even more problematic,
and not really well-defined in the absence of additional fields.)
Chatterjee and Zamolodchikov \cite{cz} show that the asymptotic
form of the bulk magnetization depends on the distance from the
boundary as $y^{-1/8}$ in the presence of any nonzero boundary field,
just as it does for fixed boundary conditions \cite{cl}.  Our
results demonstrate that this RG flow is preserved in an appropriate
form after coupling the theory to quantum gravity.

\subsection{The dual picture}
\label{sec:dual}

As discussed in Section 2, the Ising model on a random lattice
can also be formulated in the dual picture, in terms of matrices $X$
and $Y$, where $X$ denotes an edge separating two equal spins and
$Y$ an edge separating two opposite spins. We have seen in the
introduction that the change of variables
\begin{equation}
\begin{array}{rcl}
U &\to& {1\over \sqrt{2}}(X+Y)
\\
V &\to& {1\over \sqrt{2}}(X-Y)
\end{array}
\end{equation}
in the action for the Ising model leads to the dual action. Thus any
calculations in the original model can be reinterpreted in terms of
the dual model. We shall now discuss the duals of our results for the
boundary and bulk magnetizations in the presence of a boundary
magnetic field.

First, let us discuss the boundary conditions corresponding to the
weights 
\begin{equation}
\langle \Tr\; Q^n \rangle = \langle \Tr\; (e^h(X+Y) + e^{-h}(X-Y))^n\rangle
= \langle \Tr\; (\cosh(h)\, X +
\sinh(h)\, Y)^n\rangle
\end{equation}
where we have dropped factors of $\sqrt{2}$ for notational simplicity.
In the dual variables, $h=0$ corresponds to fixed boundary conditions,
while $h=\pm\infty$ corresponds to the two types of free boundary
conditions, $(X\pm Y)^n$.  Thus, in this picture $h$ plays the role of
a ``boundary freedom field'' rather than a boundary magnetic field. In
the limit $h\to 0$ the boundary condition is fixed (all $X$'s), while
in the limits $h\to \pm \infty$, the boundary conditions are a pair of
free boundary conditions with different signs in the weights assigned
to configurations with an odd number of $Y$'s on the boundary. These
fixed and free boundary conditions in the dual model are of course
precisely the Kramers-Wannier duals of the free and fixed boundary
conditions of the spin representation.  Correspondingly \cite{cardy},
a spin operator in the original variables is transformed into a
disorder operator in the dual variables.


\begin{figure}
\centerline{
\psfig{figure=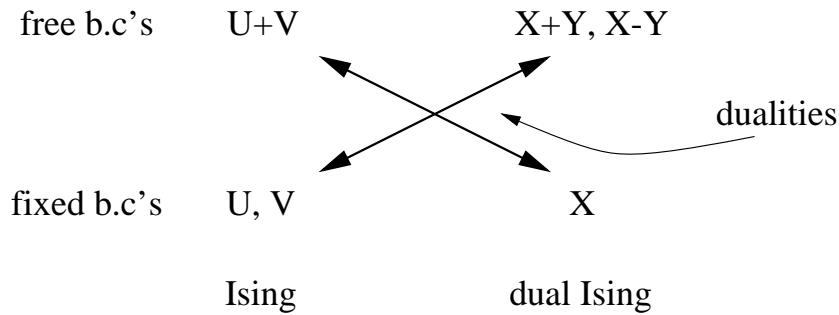,angle=-90,height=4cm}}
\caption{The duality map that interchanges free and fixed boundary
conditions. Note that for the Ising model, there are two fixed states
and only one free state, whereas the opposite is true of the dual model.}
\end{figure}

Our results in Sec. \ref{sec:bomf} for the boundary magnetization can 
thus be reinterpreted as a calculation of the expectation value of the
boundary disorder operator. We see that as the boundary freedom field
is increased, so the expectation value of the boundary disorder operator
\begin{eqnarray}
  \langle d \rangle_{n,k} &=&
  {\langle \Tr\; (e^h(X+Y) - e^{-h}(X-Y))(e^h(X+Y) + e^{-h}(X-Y))^k\rangle
  \over \langle \Tr\; (e^h(X+Y) + e^{-h}(X-Y))^{k+1}\rangle}
  \nonumber\\
  &=&
  {\langle\Tr\;( \sinh(h)\, X + 
 \cosh(h)\, Y)
  ( \cosh(h)\, X + 
  \sinh(h)\, Y)^{k}\rangle_n\over\langle\Tr\;
  ( \cosh(h)\, X + 
  \sinh(h)\, Y)^{k+1}\rangle_n}\ 
  \label{eq:bd}
\end{eqnarray}
tends to $1$. At $h=0$ the
boundary condition is fixed and so, by symmetry, the
boundary disorder must vanish.

The dualization of the calculation of the bulk magnetization goes
along similar lines, and we conclude that the bulk disorder in the
presence of a boundary freedom field is given by the expression
\begin{equation}
  \langle D \rangle=
  L^{1/3}A^{-1/3}
  \label{eq:bulkdisorder}
\end{equation}
except when $d=0$, when the bulk disorder also vanishes identically.
There is a striking consequence of this result. In the Ising model in
$U$, $V$ variables, the coupling of a boundary magnetic field was seen
to induce renormalization group flow from free to fixed boundary
conditions. On the other hand, when a boundary freedom field is
coupled in the dual formulation of the model, the renormalization
group flow is from fixed to free boundary conditions. This change in
the direction of the renormalization group flow is a natural
consequence of the Kramers-Wannier duality of the system.  The duality
symmetry implies that the ground state degeneracies of the free and
fixed states are swapped under the duality transformation, and so the
reversal of RG flow is consistent with Affleck and Ludwig's g-theorem
\cite{al}.  In general, one expects that the RG flow of the theory
should be towards the conformal boundary condition with smaller
degeneracy, so it is natural that the direction of the flow switches
under the duality transformation.  As we see, this result seems to
hold in the theory equally well after coupling to quantum gravity.

\subsection{The effects of gravity}

On any fixed lattice, the introduction of an external magnetic field
on the boundary leads to a direct effect on the Ising spins, as 
there are no other degrees of freedom with which to interact.  It
is therefore natural to expect that coupling to quantum gravity,
which introduces the local geometry as an additional degree of
freedom, will lead to quantitative changes in the response of the
spins to the boundary field, and indeed this is what we have
observed.

The calculation of the boundary magnetization in Section 6 can be
compared with the results that have been obtained in flat space by
McCoy and Wu \cite{mw}. They find that the magnetization scales as
$h\ln h$ for small $h$, and as a result, the magnetic susceptibility
$\chi$ diverges at the critical temperature. On the other hand, we
have seen in (\ref{eq:ss}) that the magnetic susceptibility at the
critical temperature is finite when the Ising model is defined on a
random lattice. It seems likely that the
exact numerical value (\ref{eq:ss}) of the magnetic susceptibility
$\chi$ depends on the discretization scheme we have used, and is not
universal.  However, we expect the fact that $\chi$ is a finite
constant to be universal (independent of the specific discretization
scheme chosen), although in the absence of calculations in alternative
schemes, this remains a conjecture.

The effect of gravity is therefore to soften the
initial impact of the boundary field.  It is natural to interpret
this softening as being due to the interaction between the
spins and the geometry; the coupling between spins and the
boundary field changes the relative weighting of different 
geometries, which changes in turn the effect of the neighboring
spins on any one boundary site, leading to a more gradual increase
in the boundary magnetization as a function of boundary field.

A related aspect of our results is that the bulk magnetization
is seen to decrease when a nonzero boundary magnetic field is increased 
(at least asymptotically, for large areas).  At
first this seems implausible, and indeed at a fixed point in flat
space, the bulk magnetization cannot behave in this way. 
However, the sum over geometries provides a possible explanation
for this unusual effect.  The expectation value of a spin in the
bulk naturally depends not only on the magnitude of the boundary
field, but also on the average distance of the point from the
boundary.  Therefore, the decrease in the bulk magnetization can
arise if the boundary field alters the relative weights of 
different geometries in such a way as to move a typical interior
point further away from the boundary (as in Figure 5).

\begin{figure}
\centerline{
\psfig{figure=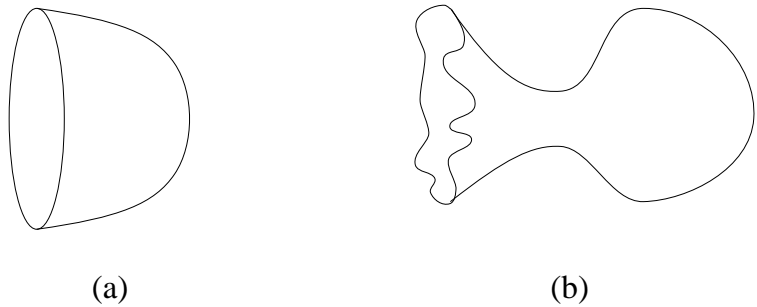,angle=-90,height=4cm}}
\caption{The relative weight of geometries of different shapes changes
  with boundary magnetic field $h$. Roughly speaking, for
  large $h$, geometries of type (a) are suppressed relative to type
  (b).}
\end{figure}

We have therefore verified that a number of features of the Ising
model in flat space are maintained in the presence of quantum 
gravity, while also demonstrating that the dynamical geometry does
have a measurable effect.  It would be interesting to check more
directly that the explanations we have given for these phenomena
are correct, for example by numerical simulation methods such as those
described recently in \cite{numeric}.

\section*{Acknowledgements}

We would like to thank D.\ Abraham, V.\ Kazakov,  J.\ Polchinski,
L.\ Thorlacius and G.\ Watts for helpful conversations.
This work was supported in part by the National Science Foundation
under grants PHY/94-07195 and PHY96-00258 and by the PPARC, UK.


\begin{thebibliography}{999}
\parindent=.6em

\bibitem{kazakov} V.\ Kazakov, {\sl Phys.\ Lett.} {\bf 119A} (1987) 140;
D.\ V.\ Boulatov and V.\ A.\ Kazakov, {\sl Phys.\ Lett.}
{\bf 186B} (1987) 379.

\bibitem{cot1} S.\ M.\ Carroll, M.\ E.\ Ortiz and W.\ Taylor, 
{\sl Nucl. Phys.} {\bf B468} (1996) 383, {\tt hep-th/ 9510199}.

\bibitem{cot2} S.\ M.\ Carroll, M.\ E.\ Ortiz and W.\ Taylor,
{\sl Nucl. Phys.} {\bf B468} (1996) 420, {\tt hep-th/ 9510208}.

\bibitem{staudacher} M.\ Staudacher, {\sl Phys. Lett.} {\bf B305} (1993) 332,
{\tt hep-th/9301038}.

\bibitem{dl}M.\ R.\ Douglas and M.\ Li, {\sl Phys.\ Lett.} {\bf B348} (1995)
360, {\tt hep-th/9412203}.

\bibitem{cot3} S.\ M.\ Carroll, M.\ E.\ Ortiz and W.\ Taylor,
{\sl Phys. Rev. Lett.} {\bf 77} (1997) 3947, {\tt hep-th/9605169}.

\bibitem{mw}  B.\ M.\ McCoy and T.\ T.\ Wu, {\em Phys.\ Rev.} {\bf
162},  436 (1967); {\em Phys.\ Rev.} {\bf
174},  546 (1968); {\em The two-dimensional Ising model} (Harvard
University Press, 1973).

\bibitem{gz} S.\ Ghoshal and A.\ Zamolodchikov, {\sl Int.\ J.\ Mod.\
Phys.} {\bf A9} (1994) 3841, {\tt hep-th/ 9306002}; Erratum: {\bf A9}, 4353.

\bibitem{cz} R.\ Chatterjee and A.\ Zamolodchikov, {\sl Mod.\ Phys.\ 
Lett.} {\bf A9} (1994) 2227, {\tt hep-th/ 9311165}.

\bibitem{klm} R.\ Konik, A.\ LeClair and G.\ Mussardo, {\sl Int.\ J.\ Mod.\ 
Phys.} {\bf A11} (1996) 276, {\tt hep-th/9508099}.

\bibitem{cardy} J.\ Cardy, {\sl Nucl.\ Phys.} {\bf B324} (1989) 581.

\bibitem{al}  I.\ Affleck and A.\ Ludwig, {\sl Phys.\ Rev.\ Lett.} {\bf
67}, 161 (1991).

\bibitem{polchinski} J.\ Polchinski, {\it TASI lectures on
D-branes}, {\tt  hep-th/9611050}.

\bibitem{kostov} I.\ Kostov, {\sl Mod.\  Phys.\  Lett.} {\bf A4} (1989) 217.

\bibitem{sy} F.\ Sugino and T.\ Yoneya, {\sl Phys. Rev.},
{\bf D53} (1996) 4448, {\tt hep-th/9510137}.

\bibitem{asatani} T.\ Asatani, T.\ Kuroki, Y.\ Okawa, F.\ Sugino, and
T.\ Yoneya, {\sl Phys. Rev.}, {\bf D55} (1997) 5083, {\tt hep-th/9607218}.

\bibitem{kuroki} T.\ Kuroki, Y.\ Okawa, F.\ Sugino, and T.\ Yoneya,
{\sl Phys. Rev.}, {\bf D55} (1996) 6429,
{\tt hep-th/9611207}.

\bibitem{siegel} W. Siegel, {\sl Phys. Rev.} {\bf D54} (1996) 2797,
  {\tt hep-th/9603030}.

\bibitem{mss}  G. Moore, N. Seiberg, and M. Staudacher, {\sl Nucl.
Phys.} {\bf B362} (1991) 665.

\bibitem{gn} E. Gava and K. S. Narain, {\sl Phys. Lett.} {\bf 263B} (1991)
213.


\bibitem{kpz} V. G. Knizhnik, A. M. Polyakov, and A. B.
Zamolodchikov, {\sl Mod.  Phys.  Lett.} {\bf A3} (1988) 819.

\bibitem{ddk}F. David, {\sl Mod. Phys. Lett.} {\bf A3} (1988) 1651; 
J. Distler and H. Kawai, {\sl Nucl. Phys.} {\bf
B321} (1989) 509.

\bibitem{cl} J.\ Cardy and D.\ Lewellen, {\sl Phys. Lett.} {\bf 259B} (1991)
274.

\bibitem{numeric} J.\ Ambjorn, K.\ N.\ Anagnostopoulos, U.\ Magnea and
G.\ Thorleifsson, {\sl Nucl. Phys. Proc. Suppl} {\bf  53} (1997) 725,
{\tt hep-lat/9608022}; J.\ Ambjorn, K.\ N.\ Anagnostopoulos and U.\
Magnea, {\sl  Mod. Phys. Lett.} {\bf A12} (1997)  1605,
{\tt hep-lat/9705004}; M.\ Bowick, V.\ John and G.\ Thorleifsson, 
{\sl Phys.  Lett.} {\bf B403} (1997) 197, {\tt  hep-th/9608030}.

\end{thebibliography}
\end{document}